# Non-fungible Tokens: Promise or Peril?

Arsalan Parham
Data and Knowledge Engineering Faculty
Bergische Universität Wuppertal
Wuppertal, Germany
arsalan.parham@uni-wuppertal.de

Corinna Breitinger
Data and Knowledge Engineering Faculty
Bergische Universität Wuppertal
Wuppertal, Germany
corinna.breitinger@uni-wuppertal.de

## ABSTRACT

Non-fungible tokens or NFTs are the digital assets on a blockchain. NFTs are unique and they cannot be divided like cryptocurrencies. NFTs could store digital ownership of artwork or collections or can be fan tokens or tickets for clubs. NFTs are based on a smart contract on a blockchain network which supports them such as Ethereum, Cardano, or Polkadot. Most of the NFTs are now minted on Ethereum (ERC-20) network but it has some main issues like high transaction fees and low speed. There are lots of domains which can be benefited from NFT technology such as art, music, gaming, sport, and wildlife conservation. NFTs could also be bought or sold on many NFT marketplaces such as Opensea and Chiliz. The trend is in huge hype because the market cap and popularity of NFTs are growing significantly.

## KEYWORDS

NFT, non-fungible tokens, Ethereum, fan tokens

## 1 Introduction

NFTs are now one of the important parts of a blockchain and decentralized technology that are trending in many different industries in the world. NFTs are implemented on blockchain networks.

NFT was firstly proposed in Ethereum blockchain (EIP-721) and it is derived by smart contracts of Ethereum. NFTs differ from cryptocurrencies like Bitcoin. NFTs are non-fungible and unique tokens that cannot be divided like digital coins. NFTs represent the existence and ownership of digital assets such as videos, images, tickets, artworks, etc [23].

The first idea of NFTs originated from "Colored Coins" in 2012 by Yoni Assia. Colored Coins or Bitcoin 2.x are unique and identifiable units of a Bitcoin that can be small as a single Satoshi (0.00000001 BTC). Colored Coins' transactions are identifiable from regular bitcoin transactions and they can represent a multitude of assets and have multiple use cases [29].

There are lots of different blockchain networks but which of them are suitable for implementing the NFTs?

The most used blockchain network that now supports many NFT projects is Ethereum. Ethereum can also provide smart contracts which are required to mint NFTs. But Ethereum is not the best in all aspects and it has several significant issues. We analyze the challenges and issues like high transaction fees and low speed that are the main problems of Ethereum blockchain. This leads to using other blockchain networks or the need to upgrade its protocols. But which other alternative blockchains are available and can be used for NFTs instead of Ethereum?

There are several alternative blockchain networks which support the architecture of NFTs like Cardano, Polkadot, or Ethereum 2.0. These networks could solve the problems of Ethereum blockchain and they can be new alternative networks or competitors for Ethereum in the future. We compare them in several aspects like speed and fees.

NFTs have useful applications in technology and numerous domains such as art, sport, music, video games, or even wild conservation. Therefore, NFTs are being very popular amount people from outside of the technology like artists, gamers, DJs or football fans could make a great future and huge market cap for NFTs.

In this paper, we first introduce basic concepts about blockchain technology and the general definition of NFT. Then, several different blockchain networks which support NFTs and their standards, differences, advantages, and disadvantages will be compared and introduced in detail. Next, we survey briefly the smart contracts and solidity programming language that are one of the important requirements of NFT's implementation. Then, we answer the question "What are the use cases of NFTs?". In addition, several projects for each domain of NFTs are introduced. In the challenges part, we analyze the main issues and challenges of the NFTs and give potential solutions to tackle the problems. Finally, we discover the story of the NFTs market and changes in the last few years and whether there is a correlation between the market price of cryptocurrencies such as Bitcoin and NFTs' price.

## 2 Background

Satoshi Nakamoto introduced the first decentralized peer-to-peer electronic cash system and blockchain network [1]. Then, the world of blockchain and decentralized technologies rose and lots of other new-generation blockchain networks were implemented with new technologies in speed, consensus algorithm, and smart contract implementations. A blockchain is the heart of NFTs technology because the main target of NFTs and blockchain technologies is to implement and use a decentralized network instead of traditional basic central ecosystems.

## 2.1 What is a blockchain

Blockchain is a decentralized network that is initiated by nodes that create a new "block" of data. A blockchain node is an open-source, cross-platform runtime that allows developers to create various services [28]. Blockchain uses an encryption method called cryptography. All the new blocks are broadcasted to every party over the network in an encrypted form and they are signed with a digital signature called private key [10].

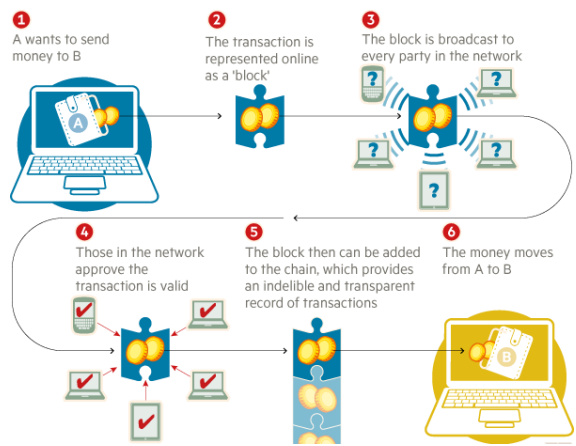

Figure 1: How a blockchain works [22]

## 2.2 NFTs vs. fungible Tokens

There are several main differences between NFT and fungible tokens. Fungible tokens like Bitcoin and fiat currencies are identical or uniform and can be interchanged with other fungible tokens of the same type without any issues.

NFTs have two major differences from fungible tokens. NFTs cannot be divided or merged and they are unique. NFT stands for representing the digital ownership of digital collections, artworks, or any other digital certification of ownership of any digital asset in a blockchain network [8]. The first NFT was introduced on the Ethereum ERC-721 blockchain in 2017. In comparison to fungible tokens, NFTs have several new use cases such as representing game collectible items, fan tokens, and digital artworks [2].

# 3 Blockchains for NFTs

## 3.1 Ethereum

The Ethereum blockchain is the second biggest blockchain network and also supports the implementation of NFT. Ethereum was proposed by Vitalik Buterin in 2013 [5], because of the lack of smart contracts in Bitcoin's blockchain network. This decentralized programmable blockchain provides smart contracts which are used by lots of NFT projects.

### 3.1.1 ERC standards

Several standards are accepted on the Ethereum blockchain for different use cases. The tokens are implemented in the solidity programming language.

**ERC-20**: A general standard for tokens which is mostly used on the Ethereum blockchain with basic functionality to transfer tokens through a smart contract.

**ERC-721**: A common non-fungible standard that each token is a unique distinguishable asset. With this standard, each NFT has its ownership in a smart contract [3].

**ERC-821**: Through this standard, anyone can check the address of the contract and find which features are supported. There are also other standards for NFTs on blockchains such as NEO and EOS [4].

**ERC-1155**: A new standard that supports the features of fungible (ERC-20) and non-fungible tokens (ERC-721). It also supports converting and minting new tokens [21].

### 3.1.2 Ethereum 2.0

Ethereum 2.0 is the second official version of the Ethereum network which solves the main issues of current Ethereum blockchain problems in several phases.

The most important issue of the current Ethereum network is scalability. There are a high number of decentralized applications of the Ethereum network that process approximately up to 20 transactions per second. It leads to a lower speed and higher costs in the Ethereum network. Ethereum 2.0 will allow performing a thousand transactions per second, to solve the scalability problem [17].

This network upgrade will also change the current consensus protocol. It transmits the current proof-of-work model to proof-of-stake which makes the Ethereum network more environmentally friendly because of lower energy consumption [17]. Ethereum 2.0 is already launched on 1 December 2020 and the last phase will be launched probably in 2021.

## 3.2 Cardano

Cardano is one of the popular developing third-generation blockchain platforms which was started by Charles Hoskinson, the former co-founder of Ethereum [9] after he left the Ethereum project in 2014.

Cardano is open-source and has several advantages against Ethereum and Bitcoin such as higher speed and lower transaction fees. It supports decentralized applications (so-called "Dapps") and smart contracts that provide minting the NFTs. Another interesting feature of the Cardano project is that it is one of the first blockchain projects which is developed and designed by a scientific team of academics [10].

IOHK company was one of the important developers of the Cardano project. IOHK has a team of engineers and scientists that builds cryptocurrencies and blockchains for corporations and academic institutions [14].

Cardano uses the proof-of-stake [PoS] protocol which has several advantages over the proof-of-work [PoW] protocol used by blockchains such as Bitcoin and Ethereum. The advantages of PoS are security, more decentralization, energy, and cost efficiency [11].

Cardano is the most staked network in the world with about 73% of all the network coins [12]. It means that Cardano, Polkadot, and Ethereum 2.0 blockchains have the most potential capacities to become the future of blockchain technology as the staked values.

## 3.3 Polkadot

Polkadot is a next-generation blockchain protocol that unites and connects parachains, a network of heterogeneous blockchain shards through the Polkadot Relay Chain [16].

Polkadot is also capable of supporting NFTs in its ecosystem. Polkadot has several advantages over the Ethereum network such as lower transaction fees and higher speed. Therefore, Polkadot can be an alternative ecosystem for NFTs. Polkadot has a wide range of partners and chains which are adopted to its technology [16].

| Network | Transaction fee | Speed | Consensus algorithms |
|---|---|---|---|
| *Ethereum* | ~ 10 - 40 $ | 15 - 20 Tps | PoW |
| *Ethereum 2.0* | ~ 1 - 8 $ | 100,000 Tps | PoS |
| *Cardano* | ~ 0.3 $ | 1,000,000 Tps | PoS |
| *Polkadot* | ~ 2 $ | 166,666 Tps | PoS |

Table 1: Compare common blockchain networks for NFTs

## 4 Smart Contracts

To mint or trade a NFT on a blockchain, it is required to use smart contracts. As a technical view, "smart contracts are digital contracts allowing terms contingent on the decentralized consensus that is tamper-proof and typically self-enforcing through automated execution" [13].

A smart contract includes several lines of code and it is a self-executing digital contract with terms of the agreement that are stored in the blockchain.

## 4.1 Solidity programming language

To implement smart contracts in blockchain, we should often use Solidity programming language. Solidity is a high-level JavaScript-like programming language that is mostly used to write smart contracts on many blockchains like Ethereum. The smart contracts are similar to object-oriented programming languages and consist of functions and variables in Solidity [19].

```
pragma solidity ^0.4.17;                              1
contract SimpleDeposit {                              2
  mapping (address => uint) balances;                 3
                                                      4
  event LogDepositMade(address from, uint amount);    5
                                                      6
  modifier minAmount(uint amount) {                   7
    require(msg.value >= amount);                     8
    _;                                                9
  }                                                   10
                                                      11
  function SimpleDeposit() public payable {           12
    balances[msg.sender] = msg.value;                 13
                                                      14
                                                      15
  function deposit() public payable minAmount(1 ether) 16
    balances[msg.sender] += msg.value;                17
    LogDepositMade(msg.sender, msg.value);            18
                                                      19
                                                      20
  function getBalance() public view returns (uint     21
    balance) {
    return balances[msg.sender];                      22
                                                      23
                                                      24
  function withdraw(uint amount) public {             25
    if (balances[msg.sender] >= amount) {             26
      balances[msg.sender] -= amount;                 27
      msg.sender.transfer(amount);                    28
    }                                                 29
  }                                                   30
}                                                     31
```

Figure 2: a simple smart contract example in solidity programming language [19]

The Ethereum platform and Solidity are evolving rapidly to avoid bugs and security risks and the developers should consider these developments. For example, the code compiled today may not be compiled a few months later [19].

# 5   Use cases of NFTs

There are several main domains of NFTs that have the biggest share of the NFTs' market. The most-traded NFTs are in the categories of art, collectible objects, and games collectibles by about 85% of total traded volume.

There are also new trending use cases of NFTs in sports like fan tokens that are being popular among football clubs or in other sports.

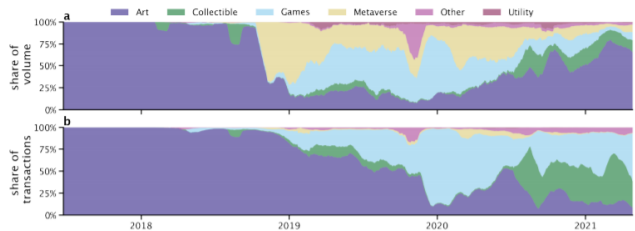

Figure 3: Composition of the NFT market. (a) Share of volume traded by category. (b) Share of transactions by category [18].

## 5.1   Artworks

One of the first and biggest domains of NFTs is Artworks. Artists can create an NFT from their artworks and sell it as a digital asset or digital ownership on a blockchain network. As figure 3, art is now the biggest share of the NFTs market cap and has an important role in the NFTs world.

Art NFTs could include different types of files such as digital images, gifs, video files, or other intellectual properties. The artists had lots of challenges to display their works and get attention. They had to invest lots of money in advertisements or pay fees by traditional platforms. Therefore, artists cannot sell their artworks as their real value of them. With the idea of NFTs, the artists do not need any agents to certify the ownership of any assets [23].

There are several secure NFT marketplace platforms that artists can represent their artworks or sell them to any other person with a low fee.

### 5.1.2   Opensea

Opensea[1] is a decentralized marketplace for buying, selling, and trading through smart contracts on the Ethereum blockchain. The users store their items in their chosen wallet like Metamask wallet. Opensea allows the users to trade not only artworks but also game collections, sport NFTs, domain names, and digital collectibles by only a 2.5% purchasing fee that making it one of the most affordable platforms for NFTs.

## 5.2   Music

Like art, the music industry is affected by NFT and blockchain technology. Every piece of music can be converted to a unique NFT. Singers or DJs can sell limited copies of any music which they produced. This use case of NFTs has several benefits for both artists and their fans. The NFTs could also be traded between the fans and they can become financially the stakeholders of the music industry.

### 5.2.1   Audius

Audius[2] is a vast developing NFT project in the music industry and it is known as Spotify killer. With the decentralized technology, the artists can earn more money from their works in comparison with traditional ways. For example, only 12% of the whole revenue generated in the music industry is earned by the artists in 2017. The main aims of the Audius project are transparency, lack of laws, and in-time payments to artists. It allows high-quality audio streaming for everyone in a decentralized network. The Audius protocol is based on its token called $AUDIUS and is implemented on the Ethereum ecosystem [24].

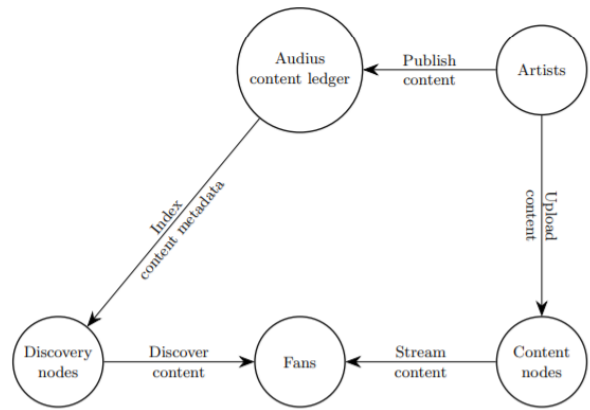

Figure 4: Audius content lifecycle [24]

Audius is well known as a rival against Spotify or Apple music streaming services. The decentralized technology of Audius is the main feature that could compete with Spotify and Apple music in the future.

## 5.3   Game collections

The game industry is another important use case of NFTs. There are a lot of game networks that have a pool of players. Players are directly connected and send data through a pool of proxies regarding their states in the game. The rules of the game are stored in the smart contracts based on the ERC-1155

---

smart contract by the creators of the game. Players could receive objects in the game and these objects or game collections can be added to the Ethereum wallets of the players. These tokens can also be traded or attached to other tokens based on the ERC-1155 standard. ERC-115 enables game objects and multiple tokens to be included in a single smart contract. All of the transactions of the players are recorded on the blockchain network. Gaming networks could be attacked or cheated, but a fail-proof monitoring mechanism can protect the network [21].

## 5.4 Fan Tokens

One of the new interesting use cases of tokens in sport is fan tokens offering. Any fan of a sports club can buy a fan token of a certain club in an exchange by paying real money. Fan token is like a stock of a club and any fan can hold a small share of it. The price of the token is dependent on such factors as the value, profit, or the popularity of a club which changes over time with any failure or success in that club [20].

The fans could also more get in touch with the club with their tokens. For instance, any token owner has the right to vote regarding some decisions about the club. This leads to a deeper connection between fans and the clubs because the clubs can better understand the taste of their fans. The fan token offering is a win-win game for both of the fans and clubs and both can benefit from it [20].

### 5.4.1 Chiliz

The Socios[3] app with its cryptocurrency called Chiliz[4] ($CHZ) is the biggest platform based on fan tokens. Chiliz has a lot of partners from high-ranked football clubs such as FC Barcelona, Juventus, Paris Saint-Germain, and Atletico de Madrid. Several other famous football clubs will join it in the future. The Chiliz market is growing rapidly and fan tokens are more popular in Asia, where crypto technology is more popular and fan bases are growing [20].

## 5.5 Wildlife conservation

NFTs have such interesting positive effects such as conserving wildlife. NFTs could store digital collection assets from lots of endangered animals and avoid extinction of wildlife and clone animals for tourism by implementing the crypto-wildlife. For example, Cryptokitties[5] is a blockchain-based collectible game through the idea of the extensibility of NFTs. Extensibility is one of the special features of NFTs which make it possible that

NFTs could be combined to create another unique NFT. The extensibility feature of NFTs allows making revenue to conserve the wildlife by running campaigns on the platforms such as Cryptokitties and Panda Earth [4].

| Category | Description | Projects | Network |
|---|---|---|---|
| **Artworks** | NFTs of digital artworks, such as images, videos, or gifs | Opensea | Polygon |
| **Music** | Audio streaming tokens | Audius, Rocki, Mozik | Ethereum, Solana |
| **Game collections** | NFTs used for game collection objects | Cryptokitties | Ethereum |
| **Fan tokens** | Tokens used for football clubs | Chiliz | Ethereum |
| **Wildlife conservation** | Wildlife NFTs | Cryptokitties, Panda Earth | Ethereum |

Table 2: Main use cases of NFTs

# 6 Challenges of NFTs

## 6.1 Gas fee

There is a dynamic fee for making any transactions on the Ethereum network called "gas fee".

The gas fee is received by the miners who validate the transactions on the blockchain. It is sent by the sender to the address of the miner as the price of the transaction. It is a common method on many blockchains called "proof of work". The Gas fee is measured by "gwei". One gwei is 0.000000001 ETH [6]. Gas fee prices are dynamic and it depends on the congestion of the network and the price of Ethereum. The lowest gas fee price is usually on Weekends when the gas prices are the lowest as the figure below shows.

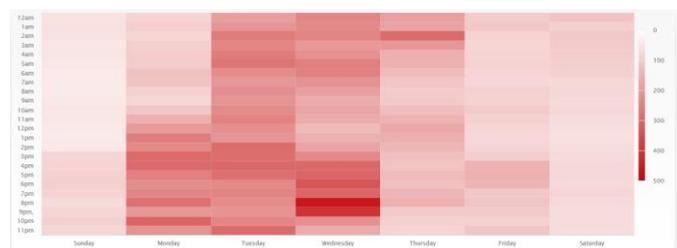

Figure 5: Gas Price by Time of Day [7]

Ethereum dynamic gas fee price is an important challenge for NFTs because most of the NFTs are minted on the Ethereum blockchain. This could be a big issue especially for cheap NFTs because this high gas fee price is not affordable for them. To tackle this problem the Cardano or Polkadot blockchains or Ethereum 2.0 version in the future would be a good choice for minting the NFTs.

## 6.2  51% attack

One of the main risks of any blockchain network is that it can be attacked by attackers. 51% attack occurs when a group of miners controls more than 50% of the network's mining hash rate or computing power.  The new transactions could be prevented from confirmations or halted by the attackers [15].
More powerful blockchains like Bitcoin does not likely to be suffering from 51% attack, because it has by far the most decentralized nodes and a very high hash rate. Therefore, it is almost impossible to be attacked, because the attacker needs lots of resources and computing power to perform it.
The more decentralized a blockchain network is, the lower chance have attackers to attack it. In conclusion, for those blockchain networks like Ethereum, the chance of being attacked is very low.

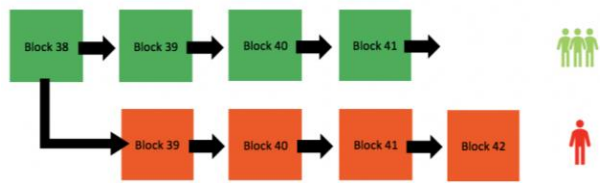

Figure 6: a 51% attack example [27]

In figure 6, the corrupt (red) miner has a higher hashing power, therefore it can add new blocks to its private chain and can broadcast its chain to the rest of the network once it is longer (heavier) than the original chain [27].

## 6.3  Extensibility issues

The extensibility feature is one of the main features of NFTs which means that NFTs can be combined to create another unique NFT.
The extensibility feature of NFTs is both an advantage and an issue. The extensibility problem is dedicated to two aspects, NFT interoperability, and updatable NFTs.
Interoperability or cross-chain means that the existing NFT ecosystems are isolated from each other and NFTs in an ecosystem can only trade within the same ecosystem or network. It is just done by cross-chain communications from external parties. Now, most of the NFTs are implemented on the Ethereum blockchain and there are no challenges regarding the extensibility for most of the NFTs. But in the future, with rising new technologies for NFTs and blockchains like Cardano and Polkadot, it can be more important than now [23].
Blockchain networks have usually several soft forks (minor modifications that are compatible forwards) or hard forks (significant modifications that may conflict with previous protocols) updates during their developments and it can cause conflicts for NFTs [23].

## 7    The market of NFTs

The NFTs' market started with rapid growth in late 2017. The majority of the market was dominated by the art category until the end of 2018, especially, by the Cryptokitties collection. From January 2019, other categories started to gain popularity. Then it remained stable until mid-2020, but from July 2020, the market experienced a dramatic growth to about 10 million US dollars daily volume in March 2021. It is about 150 times more than 8 months before [18].
The price of NFTs has almost no correlation to the cryptocurrency prices like Bitcoin because NFT markets might contain multiple asset classes. There is also little spillover between different NFT markets and it is unlike the cryptocurrencies and stock market which have high spillover among each other as the research done by [26].

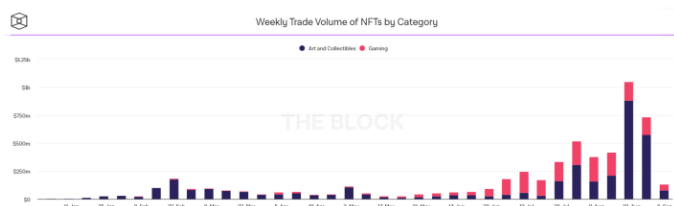

Figure 7: Weekly trade volume of NFTs by Category [30]

## 8    Conclusion

In this survey, we analyzed all the aspects of the NFTs such as architecture, domains, projects, and market in detail. The main motivation of this paper was the huge hype of NFTs and blockchain technology which has interesting functions in lots of domains in the world.
Most of the NFT projects are now using the Ethereum ecosystem, but it can change in the future because of its lack and problems. The projects with a high potential chance of being an alternative to Ethereum blockchain are Cardano, Polkadot, and Ethereum 2.0.
NFTs can make great changes in art, sport, music, gaming, or even wildlife conservation. For example, the main share of ticket offerings by football clubs can be offered through fan tokens in the next years. Also, decentralized music streaming projects can be more popular among artists because of higher earnings than current traditional ways in the music industry. There are several developing projects in this field based on NFT and decentralized technology, but it takes time to be popular among people and industries in the world.

The NFT technology tends to continue its development and popularity in many areas in the future. NFT is a growing trend in crypto and blockchain technology and has only a small share of the whole crypto technology. Experts believe that it takes time to be popular among all of the people. The NFT and decentralized technology can grow rapidly because decentralization is the technology of the future against the traditional central technology.